\documentstyle[11pt,newpasp,twoside,epsf]{article}
\def\edcomment#1{\iffalse\marginpar{\raggedright\sl#1\/}\else\relax\fi}
\reversemarginpar

\begin{document}

\title{The spin periods of magnetic cataclysmic variables}

\author{A.J. Norton, R.V. Somerscales}
\affil{Department of Physics and Astronomy, The Open University, Walton Hall,
Milton Keynes MK7 6AA, U.K.}
\author{G.A. Wynn}
\affil{Department of Physics and Astronomy, The University of Leicester,
University Road, Leicester LE1 7RH, U.K.}

\begin{abstract}
We have used a model of magnetic accretion to investigate the rotational
equilibria of magnetic cataclysmic variables (mCVs). This has enabled us to
derive a set of equilibrium spin periods as a function of orbital period and
magnetic moment which we use to estimate the magnetic moments of all known
intermediate polars. We further show how these equilibrium spin periods relate
to the polar synchronisation condition and use these results to calculate the
theoretical histogram describing the distribution of magnetic CVs as a function
of $P_{\rm spin} / P_{\rm orb}$. We demonstrate that this is in remarkable
agreement with the observed distribution assuming that the number of systems as
a function of white dwarf magnetic moment is distributed according to $N(\mu_1)
{\rm d}\mu_1 \propto \mu_1^{-2} {\rm d}\mu_1$.  
\end{abstract}

\section{The sample of mCVs}

As shown in Figure 1, the mCVs occupy a wide range of parameter space in the
spin period ($P_{\rm spin}$) versus orbital period ($P_{\rm orb}$) plane. This
indicates that accretion occurs by a variety of different types of magnetically
controlled flow.  However, certain regions of the diagram are sparsely
populated. In order to understand this distribution we use a model of magnetic
accretion to investigate the rotational equilibria of these systems. First we
discuss the systems included in Figure 1 and their characteristics.

\begin{figure}[h]
\plotfiddle{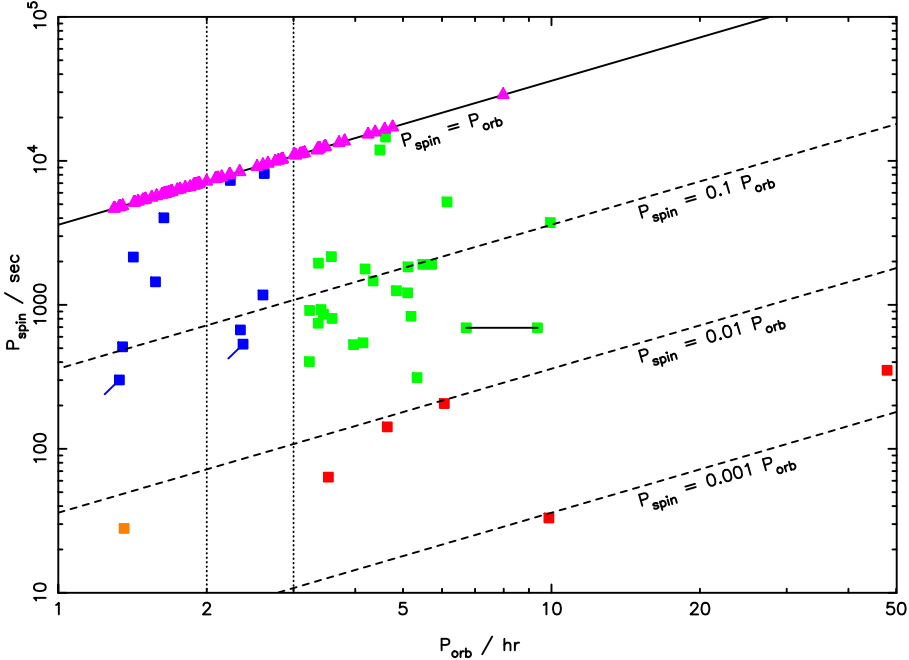}{9cm}{-90}{50}{50}{-200}{260}
\caption{The spin and orbital periods of the mCVs. Polars are indicated by 
triangles and intermediate polars by squares.}
\end{figure}

\subsection{Synchronous polars}

We include 67 synchronous polars ranging from CV Hyi with a period of 
77.8 min to V1309 Ori with a period of 7.98 hr. Magnetic field strength 
estimates exist for 52 of these, spanning the range $\mu_1 \sim 3 \times
10^{33} - 10^{35}$~G~cm$^3$. It is apparent that a true period gap no
longer exists for polars, however we note that there are 22 synchronous 
systems with $P_{\rm spin} = P_{\rm orb} >2.5$~hr and a further 45 systems 
with $P_{\rm spin} = P_{\rm orb} <2.5$~hr.

\subsection{Asynchronous polars}

There are 5 systems in which $P_{\rm spin}$ and $P_{\rm orb}$ differ by around
2\% or less. These are assumed to be polars that have been disturbed from
synchronism by a recent nova explosion. The systems are V1432 Aql with $P_{\rm
spin}/P_{\rm orb} = 1.003$ (Friedrich et al 1996), BY Cam with $P_{\rm
spin}/P_{\rm orb} = 0.992$ (Silber et al 1992), V1500 Cyg with 
$P_{\rm spin}/P_{\rm orb} = 0.983$ (Schmidt \& Stockman 1991),
V4633 Sgr with $P_{\rm spin}/P_{\rm orb} = 1.023$ (Lipkin et al 2001), 
and CD Ind with $P_{\rm spin}/P_{\rm orb} = 0.989$ (Schwope et al 1997). In
Figure 1 these systems are essentially indistinguishable from the synchronous
polars.

\subsection{Nearly synchronous intermediate polars}

Slightly further away from synchronism than the asynchronous polars are 4
recently discovered systems that may best be described as nearly synchronous
intermediate polars. The systems are V381 Vel with $P_{\rm spin}/P_{\rm orb} =
0.910$ (Tovmassian, these proceedings), RXJ0524+42 with $P_{\rm spin}/P_{\rm
orb} = 0.866$ (Schwarz, these proceedings), HS0922+1333 with $P_{\rm
spin}/P_{\rm orb} = 0.882$ (Tovmassian, these proceedings) and V697 Sco with
$P_{\rm spin}/P_{\rm orb} = 0.737$ (Warner \& Woudt 2002). Two of
these systems lie in the `period gap'. We suggest that all four systems are
intermediate polars in the process of attaining synchronism and evolving into
polars.

\subsection{Conventional intermediate polars}

We include 22 systems that we classify as conventional intermediate polars. 
These all have spin to orbital period ratios in the range $0.25 > P_{\rm
spin}/P_{\rm orb} > 0.01$ and $P_{\rm orb} > 3$~hr. The systems are V709 Cas
(Norton et al 1999), 1RXSJ154814.5--452845 (Haberl, Motch \& Zickgraf 2002),
2236+0052 (Woudt \& Warner, these proceedings), V405 Aur (Allan et al 1996), YY
Dra (Haswell et al 1997), PQ Gem (Duck et al 1994), V1223 Sgr (Taylor et al
1997), AO Psc (Taylor et al 1997), HZ Pup (Abbott \& Shafter 1997), UU Col
(Burwitz et al 1996), FO Aqr (de Martino et al 1999), V2400 Oph (Buckley et al
1997), BG CMi (de Martino et al 1995), TX Col (Norton et al 1997),
1WGA1958.2+3232 (Norton et al 2002), TV Col (Vrtilek et al 1996), AP Cru (Woudt
\& Warner 2002), V1062 Tau (Hellier, Beardmore \& Mukai 2002), LS Peg
(Rodriguez-Gil et al 2001), RR Cha (Woudt \& Warner 2002), RXJ0944.5+0357
(Woudt \& Warner, these proceedings), and V1425 Aql (Retter, Leibowitz \&
Kovo-Kariti 1998).  The orbital period of 1RXSJ154814.5--452845 is currently
uncertain, and the two possible values are both shown in Figure 1, connected by
a horizontal line.

\subsection{EX Hya-like systems}

There are a further 4 intermediate polars that lie below the `period gap'. 
These all have $P_{\rm spin}/P_{\rm orb} > 0.1$ and $P_{\rm orb} < 2$~hr. Given
their location on the $P_{\rm spin}$ versus $P_{\rm orb}$ diagram, we refer to
these as `EX Hya-like' systems for convenience. The systems are  HT Cam (Kemp
et al 2002), RXJ1039.7--0507 (Woudt \& Warner 2003a), V1025 Cen (Hellier, Wynn
\& Buckley 2002) and EX Hya (Allan, Hellier \& Beardmore 1998). The systems DD
Cir (Woudt \& Warner 2003b) and V795 Her (Rodriguez-Gil et al 2002) lie within
the `period gap' with $P_{\rm spin}/P_{\rm orb} \sim 0.1$ and may be included
with these systems.

\subsection{RXJ1914.4+2456 and RXJ0806.3+1527}

RXJ1914 (a.k.a. V407 Vul) and RXJ0806 are X-ray sources that each display a
single coherent periodiocity at 569s and 321s respectively. They have been
suggested to be double degenerate systems, where the modulation represents the
orbital period of either a polar, an Algol-like system or an electrically
powered star (Cropper et al 1998, these proceedings; Ramsay et al 2000, 2002a,
2002b; Marsh \& Steeghs 2002; Wu et al 2002; Israel et al 1999, 2002, these
proceedings). It has also been suggested that the systems are face-on
stream-fed intermediate polars (Norton, Haswell \& Wynn 2002), in which case
the periods observed are beat periods between the white dwarf spin and orbital
periods caused by the accretion stream flipping from one pole to the other
twice in each synodic period. For stream-fed flow, the orbital periods must be
$<2.37$~h for RXJ1914 and $<1.33$~h for RXJ0806. We include these systems in
Figure 1 under the assumption that they are intermediate polars. Their data
points are shown with `tails' indicating the upper limits on their spin and
orbital periods.

\subsection{Rapid rotators} 

Finally there are a mixed bag of 6 objects that are characterized by rapidly
rotating white dwarfs. They all lie in the region of the diagram defined by
$P_{\rm spin}/P_{\rm orb} < 0.01$, but with a large range of orbital periods. 
They include the very long outburst interval SU UMa star, WZ Sge (Patterson et
al 1998) with a proposed spin period of 28~s, the propeller system AE Aqr 
(Choi, Dotani \& Agrawal 1999) with a 33~s spin period, the `DQ Her systems' 
V533 Her (Thorstensen \& Taylor 2000) and DQ Her (Zhang et al 1995), and also 
XY Ari (Hellier, Mukai \& Beardmore 1997) and the long orbital period old 
nova GK Per (Morales-Rueda, Still \& Roche 1996).

\section{Spin equilibria}

The spin rate of a magnetic white dwarf accreting via a disc reaches 
equilibrium when the rate of angular momentum accreted by the white dwarf is 
equal to the braking effect of the magnetic torque on the disc close to its
inner edge.  The inner edge of such a disc may then be defined as the point
where the magnetic stresses exceed the viscous stresses.

Most models for this equilibrium spin period find that the inner edge of the
disc, at radius $R_{\rm in}$, is roughly equal to the magnetospheric radius
$R_{\rm mag}$ and also to the corotation radius, 
$R_{\rm co}$, that is the radius at which a particle in a Keplerian orbit
would orbit the white dwarf at the same rate as the white dwarf spins on its 
axis. This implies that systems whose white dwarfs have low surface magnetic
moments will be rapid rotators, whilst systems whose white dwarfs have high
surface magnetic moments will be slow rotators. 

In turn, this leads to the requirement for truncated accretion disc formation
of $R_{\rm in} \sim R_{\rm mag} \sim R_{\rm co} << R_{\rm circ}$ (i.e. the
circularisation radius at which material would first go into orbit around the
white dwarf).  With this hierarchy of radii, the spin and orbital periods of
the system must then satisfy $P_{\rm spin} / P_{\rm orb} << 0.1$. Whilst this
condition is clearly satisfied for the rapid rotators like DQ Her, V533 Her, XY
Ari and GK Per, it is clearly not true for all intermediate polars.  

Indeed, systems with $P_{\rm spin} / P_{\rm orb} > 0.1$ are unlikely to possess
discs, since this implies $R_{\rm in} \sim R_{\rm mag} \sim R_{\rm co} > 
R_{\rm circ}$.  King \& Wynn (1999) showed that the spin period of EX Hya can 
be understood as an equilibrium situation in which the magnetospheric and
corotation radii extend out to the inner Lagrangian point, $R_{\rm mag} \sim
R_{\rm co} \sim b$. They used simulations of the accretion flow for parameters
appropriate to EX Hya to show that a range of equilibrium spin periods are
possible, but that the maximum possible spin to orbital period ratio is $P_{\rm
spin} / P_{\rm orb} \sim 0.68$, in remarkable agreement with the periods of EX
Hya. King \& Wynn further suggested that a continuum of spin equilibria should
exist appropriate to a range of magnetic moments and orbital periods. The aim
of the work reported below was to explore this parameter space.

\section{The magnetic model}

The equation of motion of plasma in a magnetic field is
\begin{equation}
\rho \frac{ {\rm d}{\bf \it v}} { {\rm d}t} = - {\bf \nabla} P - 
{\bf \nabla}_{\perp} \left[ \frac{B^2}{8\pi} \right] + \frac{1}{R} 
\left[ \frac{B^2}{4\pi} \right] {\bf \it \hat{n}} + {\bf \nabla} \Phi_{\rm grav}
\end{equation}
If we assume the tension term is the dominant force, then the magnetic 
acceleration is given by 
\begin{equation}
a_{\rm mag} = \frac {1}{R \rho} \left[ \frac{B^2}{4\pi} \right]{\bf \it \hat{n}}
\end{equation}
This in turn may be parameterized by introducing the drag parameter $k$ via
\begin{equation}
a_{\rm mag} = -k(r,t,\rho) [\Omega_{\rm K} - \Omega_* ]
\end{equation}
where $\Omega_{\rm K}$ is the angular speed of the material orbiting the
white dwarf with Keplerian velocity and $\Omega_*$ is the angular speed
of the magnetic field lines.

Up to this point, the model is appropriate for any form of magnetic interaction.
However, if we now assume that the accertion flow takes the form of diamagnetic
blobs, we may express $k$ in terms of the magnetic timescale $t_{\rm mag}$ or 
the magnetic field strength $B$, the Alfv\'{e}n speed $c_{\rm A}$, and the blob
density $\rho$ and length $l$.
\begin{equation}
k \sim \frac{P_{\rm orb}}{t_{\rm mag}} \sim 
\frac{B^2(r,\theta)}{c_{\rm A}\rho l} 
\end{equation}
The blob density is approximated by
\begin{equation}
\rho \sim \frac{\dot{M} 4 \pi^2}{P_{\rm orb}^2 c_{\rm s}^3}
\end{equation}
and the blob length by
\begin{equation}
l \sim \frac {P_{\rm orb} c_{\rm s}}{2 \pi}
\end{equation}
where the sound speed is 
\begin{equation}
c_{\rm s} \sim \left( \frac{kT}{m_{\rm H}} \right)^{1/2}
\end{equation}
with $T \sim 3000$~K.

The model outlined above is implemented in a computer code and may be used 
to allow systems to evolve to their rotational equilibrium. Other examples of 
the use of the model are presented in King \& Wynn (1999), Hellier, Wynn \& 
Buckley (2002) and Ultchin, Regev \& Wynn (2002).

\section{Results}

The results of running the model to establish the allowed range of equilibrium
spin periods are shown in Figure 2, and Figure 3 shows some example accretion
flows corresponding to different regions of the parameter space explored. 
The model was run for 10 different orbital periods (80~m, 2~h, 3~h,
... 10~h) for each of 11 different values of the $k$ parameter. These $k$
values have been converted into the corresponding value of the white dwarf
surface magnetic moment $\mu_1$, assuming the diamagnetic blob prescription of
Equation 4. Lines on Figure 2 connect the points for each orbital period. The
line for an orbital period of 80~m essentially reproduces that obtained by King
\& Wynn appropriate to EX Hya. It should be noted that there is some latitude
in applying the conversion from $k$ to $\mu_1$ (mostly from the assumed values
of $\rho$ and $l$ describing the blobs), so the envelope of equilibrium periods
may shift laterally in the plane. Also, the calculations were performed for a
single value of white dwarf mass (0.7~M$_{\odot}$) and different values will
shift the results slightly in the vertical direction. Nonetheless it is
encouraging that the predicted equilibrium spin periods lie in the range at
which many mCVs are seen.  It should also be noted that the estimated magnetic
field strengths for the intermediate polars that display polarized emission
(e.g. PQ Gem, BG CMi, V2400 Oph and LS Peg) lie within the region covered by
the results.

\begin{figure}[h]
\plotfiddle{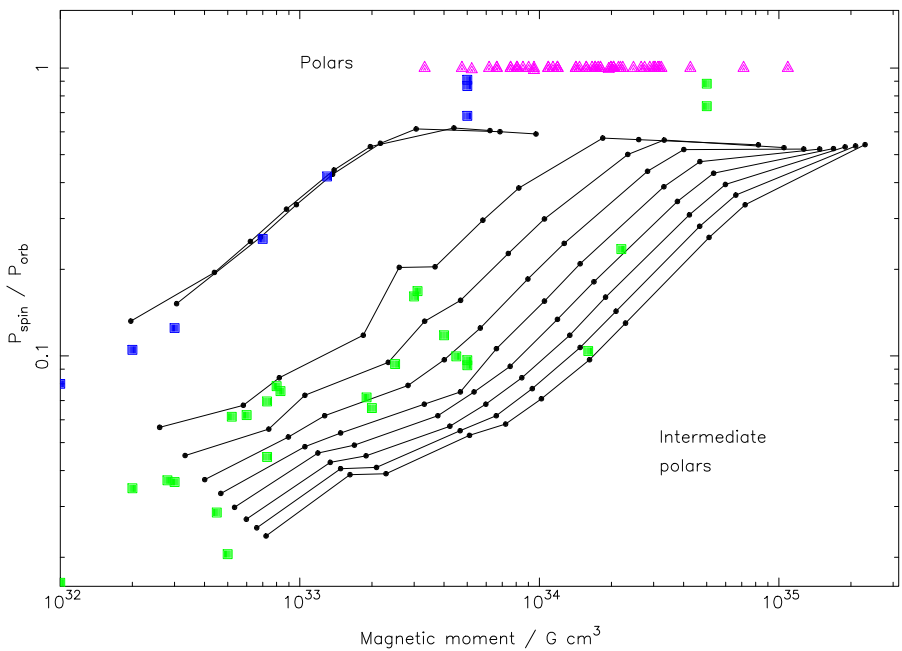}{9cm}{-90}{50}{50}{-200}{260}
\caption{The equilibrium spin periods of mCVs obtained by running the model. 
Each line connects 11 data points (11 $k$ values) and represents the 
equilibrium spin periods at a given orbital period. The ten lines correspond 
to orbital periods of 80~m, 2~h, 3~h ... 10~h. The measured magnetic moments
of polars are shown by the triangles along the top of the figure. The estimated 
magnetic moments of the intermediate polars are shown by squares and have been 
obtained by tracing across at the appropriate $P_{\rm spin}/P_{\rm orb}$ 
value to intersect the equilibrium line for the appropriate orbital period.}
\end{figure}

\begin{figure}[h]
%\plotfiddle{norton_fig03.eps}{9cm}{0}{50}{50}{-170}{0}
\plotfiddle{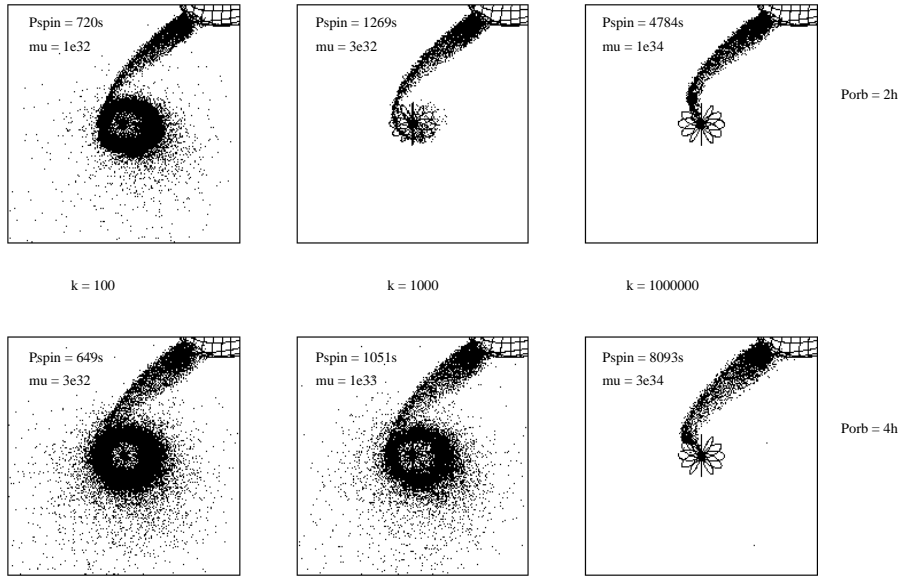}{9cm}{0}{50}{50}{-170}{0}
\caption{Examples of the accretion flows corresponding to a few of
the equilibrium situations identified in Figure 2. Truncated disc-like flows
occur for lower magnetic moments / smaller spin to orbital period ratios; 
stream-like flows occur for higher magnetic moments / larger spin to orbital 
period ratios.}
\end{figure}

In the envelope of allowed parameter space shown on Figure 2, we note that the
region with $P_{\rm spin}/P_{\rm orb} < 0.1$ corresponds to $R_{\rm mag} \sim
R_{\rm co} < R_{\rm circ}$ and so indicates regions where truncated accretion
discs may form. The region with $P_{\rm spin}/P_{\rm orb} \sim 0.1$ corresponds
to $R_{\rm mag} \sim R_{\rm co} \sim R_{\rm circ}$ and the region with $P_{\rm
spin}/P_{\rm orb} \sim 0.6$ corresponds to $R_{\rm mag} \sim R_{\rm co} \sim b$.

If we now assume that all mCVs are at their equilibrium spin period we can
place the intermediate polars on Figure 2 by tracing across at the appropriate
$P_{\rm spin}/P_{\rm orb}$ value to intersect the equilibrium line for the
appropriate orbital period. In this way, most of the squares have been placed
on Figure 2. For the few systems that have $P_{\rm spin}/P_{\rm orb} > 0.6$
(i.e. EX Hya and the 4 nearly synchronous intermediate polars) we have simply
placed them above the mid-point of the plateau part of the relevant curves at
the appropriate $P_{\rm spin}/P_{\rm orb}$ value. We note that in cases where
$\dot{P}_{\rm spin}$ is measured for intermediate polars, several are spinning
up whilst a similar number are spinning down, and FO Aqr has been seen to do
both. Hence the assumption that intermediate polars {\em are} close to their
equilibrium spin periods is a good one with only small deviations apparent on
short timescales.

\section{Synchronisation}

We now use the results obtained above to investigate the polar synchronisation
condition. We may assume that systems become synchronized once the magnetic
locking torque is equal to the accretion torque. Hence
\begin{equation}
\frac {\mu_1 \mu_2}{a^3} = \dot{M} ( G M_1 R_{\rm mag})^{1/2}
\end{equation}
where the surface magnetic moment on the secondary may be approximated by
\begin{equation}
\mu_2 = 2.8 \times 10^{33} P_{\rm orb}^{9/4}~{\rm G~cm}^3
\end{equation}
from Warner (1996) with $P_{\rm orb}$ in hours. The magnetospheric radius
is approximated by the corotation radius, 
\begin{equation}
R_{\rm mag} \sim R_{\rm co} = \left( \frac{G M_1 P_{\rm spin}^2} {4\pi^2} 
\right)^{1/3}
\end{equation}
and the mass accretion rate $\dot{M}$ is set to the secular value given by 
\begin{equation}
\dot{M} = 2.0 \times 10^{-11} 
\times P_{\rm orb}^{3.7}~{\rm M}_{\odot}~{\rm yr}^{-1}
\end{equation}
for $P_{\rm orb} > 2.7$~h (McDermott \& Taam 1989) and 
\begin{equation}
\dot{M} = 2.4 \times 10^{15} \times \frac{M_1^{2/3} P_{\rm orb}^{-1/6}}
{\left(1 - \frac{15q}{19}\right) \left(1+q\right)^{1/3} }~{\rm g~s}^{-1}
\end{equation}
for angular momemtum loss by gravitational radiation (Warner 1995) which
we assume to be dominant at $P_{\rm orb} < 2.7$~h. The mass ratio is 
$q = M_2/M_1$ where we assume a primary mass of $M_1 = 0.7$~M$_{\odot}$ and 
a secondary mass given by $M_2 = 0.091 P_{\rm orb}^{1.44}$~M$_{\odot}$.

Solutions to Equation 8 are plotted as diagonal lines on Figure 4. Each line
connects points at which synchronisation will occur for a given orbital 
period. The lines shown are for orbital periods of 3~h, 4~h, ... 10~h, as 
those for shorter orbital periods lie to the upper left of the region
plotted. Where these diagonal lines intersect the equilibrium spin period
lines for their respective orbital periods, gives the points at which systems
will synchronize. The locus of these points is shown by the thick line
which simply connects the various intersection points. Hence, systems lying
below the thick line will not be prone to synchronise, whilst those 
lying above will tend to do so.

\begin{figure}[h]
\plotfiddle{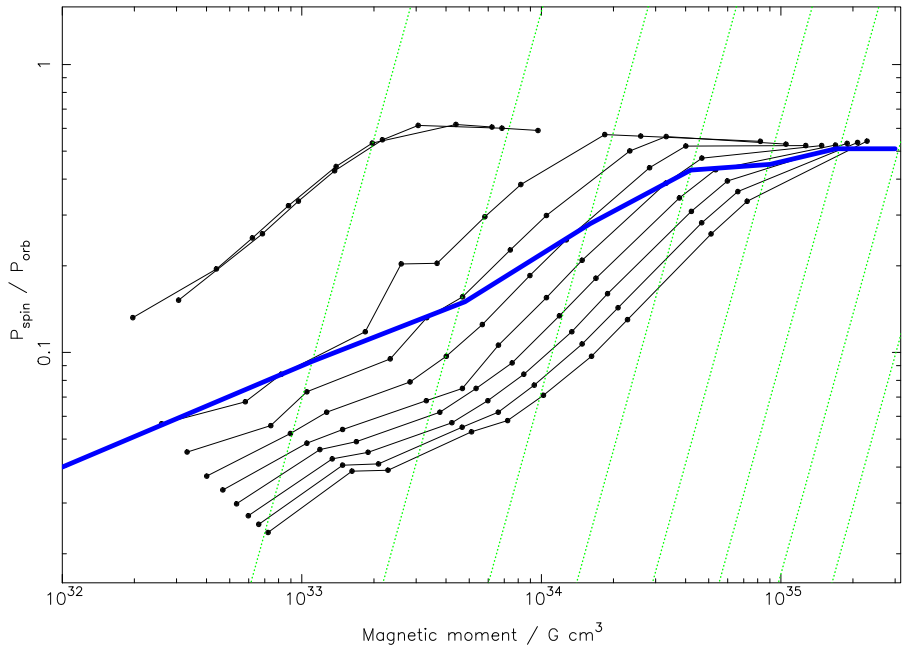}{9cm}{-90}{50}{50}{-200}{260}
\caption{Diagonal lines show solutions to Equation 5 for orbital periods
3~h, 4~h, ... 10~h. The lines for shorter orbital periods are to the upper
left of the region plotted.  Where these diagonal lines intersect the 
equilibrium spin period lines for their respective orbital periods, gives 
the points at which systems will synchronize. The locus of these points is 
shown by the thick line which simply connects the various intersection points.}
\end{figure}

\section{Predictions}

From Figures 2 and 4 it can be seen that the intermediate polars above the
period gap virtually all lie {\em below} the sychronisation line. The two
systems just above the line are RR Cha and RXJ0944.5+0357 with $P_{\rm
spin}/P_{\rm orb}$ of 0.161 and 0.168 respectively. We suggest these three are
probably normal intermediate polars and within the errors of our calculation
are consistent with sitting below the synchronisation line. Lying well above
the synchronisation line are the two nearly synchronous intermediate polars
V697 Sco and HS0922+1333 with $P_{\rm spin}/P_{\rm orb}$ of 0.737 and 0.882
respectively. We suggest that these two systems are on their way to
synchronism. A similar fate should hold for the rest of the intermediate polars
above the period gap that have relatively high white dwarf magnetic moments.

At short orbital periods, the picture is rather different. All 6 of the 
EX Hya-like intermediate polars, plus the two nearly synchronous intermediate 
polars RXJ0524+42 and V381Vel, lie above the synchronisation line and so
should be synchronised. We suggest the reason for them not being synchronised 
could be that the secondary stars in the EX Hya-like systems have low magnetic 
moments (i.e. less than predicted by Equation 9) and so are unable to come into
synchronism.

It is also apparent from Figure 2 that virtually all the intermediate polars
have $\mu_1 < 5 \times 10^{33}$~G~cm$^3$ whereas virtually all of the polars
have $\mu_1 > 5 \times 10^{33}$~G~cm$^3$. This discrepancy between the 
intermediate polar and polar magnetic fields has been noted in the past, 
and poses the question: where are the synchronous systems with low white 
dwarf magnetic moments? We can suggest three possibilities for the answer.

Firstly, it may be that the systems with a low white dwarf magnetic moment
do synchronize, but in the synchronous state they are primarily EUV emitters 
and so unobservable. Alternatively it may be that when systems with low white
dwarf magnetic moment evolve to short orbital periods, for some reason they 
simply do not synchronize, but remain on the spin equilibrium line where the
rest of the EX Hya-like systems reside. Finally there is the possibility 
raised by Cummings (these proceedings) that the magnetic fields in intermediate
polars are buried by their high accretion rate and so are not really as low
as they appear. In this picture, one could envisage that as intermediate polars
evolve towards shorter orbital periods, when they reach the period gap the
mass transfer shuts off, the magnetic field of the white dwarf re-surfaces,
and the systems synchronize before re-appearing below the period gap as 
conventional, high field, polars.

\section{The magnetic moment distribution of white dwarfs in mCVs}

The results shown in Figure 4 can also be used to characterize the theoretical
distribution of mCVs as a function of the white dwarf surface magnetic moment.
We assume that the number of systems varies according to
\begin{equation} 
N(\mu_1) {\rm d}\mu_1 \propto \mu_1^{-n} {\rm d}\mu_1
\end{equation}
where $n$ is a number to be determined. We may integrate under the 
equilibrium curves shown in Figure 4 to get the predicted number of systems
within a given range of $P_{\rm spin}/P_{\rm orb}$. The number in a given
bin is then
\begin{equation}
N \propto \left[  \frac{\mu_1^{1-n}}{1-n}  \right] ^{\log \mu_1}
_{\log \mu_1 + \frac {\Delta \log \frac{P_{\rm spin}}{P_{\rm orb}}}{G}}
\end{equation}
where $G$ is the gradient of a cubic fit to each of the equilibrium curves
in Figure 4. The range of spin to orbital period ratios was divided into 
eight logarithmic bins, seven of which were 0.25 wide in log space and
the eighth corresponded to synchronous systems. The integration
was carried out over the magnetic moment range $10^{32}~{\rm G~cm}^3 <
\mu_1 < 10^{35}~{\rm G~cm}^3$ and at the spin to
orbital period ratio (for a given orbital period) indicated by the thick line 
in Figure 4, we assume that systems become synchronized. The predicted number
of systems was averaged over orbital periods of 3, 4, 5, and 6 hours
assuming systems to be distributed evenly in orbital period, and the final 
result was normalised to the number of known mCVs with orbital periods
greater than 3~h. 

The best fit value of the power law index $n$ was found to be at 1.95, with a
reduced chi-squared of $\chi^2_r = 0.56$, as shown in Figure 5, although values
between $\sim$1.85 and 2.05 are also valid ($\chi^2_r < 1$).  The cumulative
histograms comparing the observed number of systems with that predicted by
Equation 13 with $n=1.95$ is shown in Figure 6, and Table 1.

\begin{figure}[h]
\plotfiddle{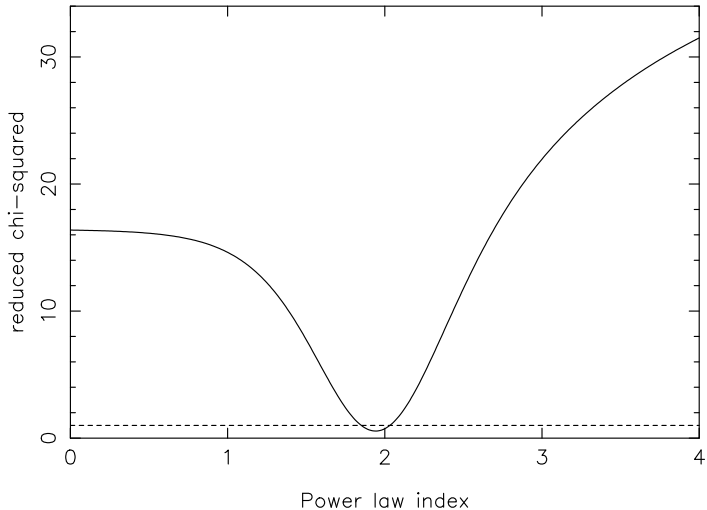}{7cm}{-90}{40}{40}{-160}{210}
\caption{The reduced chi-squared verses power law index obtained by fitting
Equation 13 to the observed number of mCVs with $P_{\rm orb}>3$~h.}
\end{figure}

\begin{table}
\caption{Cumulative histograms comparing the observed number of systems with
that predicted by Equation 13 with $n=1.95$}
\begin{center}
\begin{tabular}{cccc} \tableline
\multicolumn{2}{c}{Lower edge of bin} & & \\
log($P_{\rm spin}/P_{\rm orb}$) &  $P_{\rm spin}/P_{\rm orb}$ & $N$(obs) & $N$(model) \\ \tableline
--2.00 & 0.010 & 1  & 0.00 \\
--1.75 & 0.018 & 1  & 0.00 \\
--1.50 & 0.032 & 4  & 6.19 \\
--1.25 & 0.056 & 11 & 14.08 \\
--1.00 & 0.100 & 4  & 1.21 \\
--0.75 & 0.178 & 1  & 0.08 \\
--0.50 & 0.316 & 0  & 0.00 \\
--0.25 & 0.562 & 2  & 0.00 \\
0.00  & 1.000 & 26 & 28.45 \\
\tableline \tableline
\end{tabular}
\end{center}
\end{table}

\begin{figure}[h]
\plotfiddle{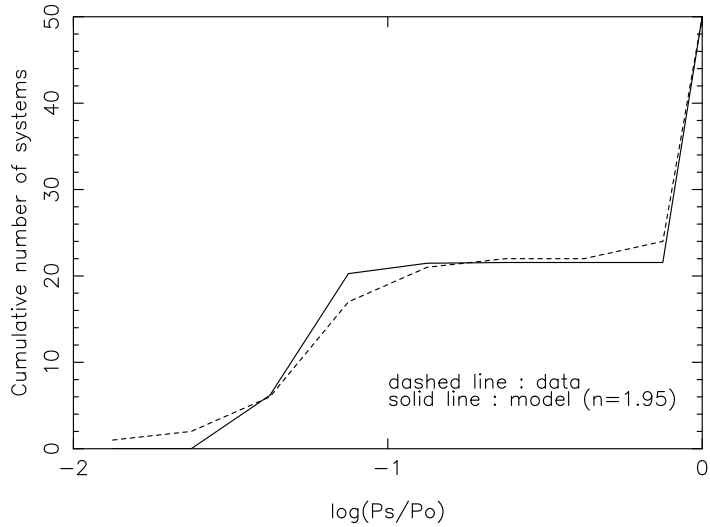}{7cm}{-90}{40}{40}{-160}{210}
\caption{Cumulative histograms showing the observed number of mCVs with 
$P_{\rm orb}>3$~h and the predicted number according to Equation 13 with 
$n=1.95$, as a function of spin to orbital period ratio.}
\end{figure}

\section{Conclusions}

We have demonstrated that there is a range of parameter space in the 
$P_{\rm spin}/P_{\rm orb}$ versus $\mu_1$ plane at which spin equilibrium
occurs for mCVs. Using the results of numerical simulations we can 
infer the surface magnetic moments of the white dwarfs in intermediate
polars to be mostly in the range $10^{32}~{\rm G~cm}^3 < \mu_1 < 5 \times 
10^{33}~{\rm G~cm}^3$. Most of the intermediate polars with orbital 
periods greater than 3~h lie below the synchronisation line, whilst most
systems with orbital periods less than 2~h lie above it and should be 
synchronized. High magnetic moment intermediate polars at long orbital
periods will evolve into polars. Low magnetic moment intermediate polars
at long orbital periods will either: evolve into EX Hya-like systems
at short orbital periods; evolve into low field strength polars which
are (presumably) unobservable, and possibly EUV emitters; or, if their 
fields are buried by high accretion rates, evolve into conventional polars
once their magnetic fields re-surface when mass accretion turns off.
EX Hya-like systems may have low magnetic field strength secondaries and 
so avoid  synchronisation.

We have also shown that the distribution of mCVs above the period gap 
follows the relationship $N(\mu_1) {\rm d}\mu_1 \propto \mu_1^{-2} 
{\rm d}\mu_1$. Recent data from G. Schmidt (private communication), concerning
the magnetic field strength distribution of isolated white dwarfs, reveals
a drop off in the number at high magnetic field strengths in those
systems too.

\end{document}